\documentclass[usenatbib]{mn2e}
\usepackage{epsfig,amsmath,amssymb,amsfonts,mathrsfs,latexsym,graphicx}
\begin{document}

\author[A.L.Watts, N.Andersson and R.L.Williams]{A.L.Watts, N.Andersson and R.L.Williams
\\School of Mathematics, University of Southampton, Southampton SO17 1BJ, UK}

\title[Differentially rotating spherical shells: The initial-value
problem]{The oscillation and stability of differentially rotating spherical shells: The initial-value
problem}

\maketitle

\begin{abstract}
An understanding of the oscillations of differentially rotating systems is key
to many areas of astrophysics.  It is of particular relevance to the
emission of gravitational waves from oscillating neutron stars, which are
expected to possess significant differential rotation immediately after
birth or binary merger.  In a previous paper we analysed the normal modes
of a simple system exhibiting differential rotation.  
In this complementary paper we address the initial value problem for the
same simple model using both analytical methods and numerical time
evolutions. We derive a necessary and sufficient condition for dynamical
shear instability.  We discuss the dynamical
behaviour of the continuous spectrum in response to an initial
perturbation, and show that certain singular solutions within the
continuous spectrum appear physically indistinguishable from the discrete
modes outside the continuous spectrum.
\end{abstract}

\begin{keywords}
hydrodynamics - instabilities - gravitational waves - stars: neutron - stars: rotation
\end{keywords}

\section{Introduction}

The oscillations of differentially rotating systems are important in many areas
of astrophysics.  They are of particular relevance to neutron star
astrophysicists, because non-axisymmetric oscillations of compact
stars are promising sources of gravitational
waves. Detection and interpretation of gravitational waves from oscillating
neutron stars could answer many questions about neutron star physics that
are difficult to answer using electromagnetic observations, for example regarding
the supranuclear equation of state.  Emission of gravitational waves
from neutron star oscillations will be strongest if the
oscillations can grow via some dynamical or secular instability. The most promising instability
scenarios arise for newly-born, accreting or newly-merged neutron stars.
But it is in precisely these situations that we expect the stars to be
differentially rotating.  The latest simulations of core collapse indicate
that the nascent neutron star emerges  rotating
differentially \citep{dm02a,aki03,ott04}.  \citet{fuj93} has shown that rapid accretion from a companion
could drive differential rotation in the surface layers of the neutron
star, and a similar effect may occur during rapid accretion of supernova
fallback material \citep{wat02}.  \citet{shi00a} and \citet{bau00} have
shown that a long-lived massive differentially 
rotating neutron star may be generated as the result of the merger of
two neutron stars. In addition, recent work on r-modes in neutron stars
suggests that the modes may drive a uniformly rotating star into
differential rotation via non-linear effects
\citep{rez00,lev01,lin01}. Understanding the effect of differential
rotation will be critical if we are to model gravitational wave emission
from these systems correctly. 
 
Differential rotation introduces two interesting phenomena.  Firstly, it may
lead to dynamical shear instabilities \citep{pap84,pap85,bal85,luy90a}.  Secondly, the
dynamical equations of the inviscid problem are complicated by the presence
of corotation points.  These are points where the pattern speed of a mode
matches the local angular velocity, and at these points the governing
equations are formally singular.  The range of frequencies for which modes
possess corotation points is termed the corotation band. It gives rise to a continuous
spectrum, the dynamical role of which is unclear.  Neither issue has
received significant attention in the stellar perturbation literature (see
\citet{wat03a}, hereafter referred to as Paper I, for a summary of previous work).  

We have begun to address
the problem by modelling a simple system, a differentially rotating incompressible
spherical shell.
This model is simplistic, but exhibits the main features of
the more complex three-dimensional problem.  In
Paper I we conducted a normal mode analysis for the
 shell. In the uniform rotation limit the shell problem admits only the standard
r-modes as solutions.  As the degree of differential rotation increases,
the frequencies of these modes gradually approach the lower edge of the
expanding corotation band.  Most modes approach the boundary
tangentially and appear to cease to exist when the degree of differential 
rotation exceeds a certain threshold value. In only a few cases do modes
cross the corotation band boundary.

For real frequencies in the corotation band the shell problem admits a
continuous spectrum of solutions with discontinuous first derivatives.  In general,
the continuous spectrum solutions have both a logarithmic singularity and a finite step in
the first derivative at the corotation point.  For certain special
frequencies, however, the step in the first derivative vanishes. We
refer to these solutions as zero-step solutions.  Since these solutions appear when 
regular normal modes cross into the corotation band, they are likely to 
be of particular importance. Furthermore, in differentially rotating cylinders,
these solutions can merge to give rise to dynamical instabilities \citep{bal85}.
Even though we do not observe such behaviour in our problem it is an 
interesting possibility that should not be overlooked in future studies. 

The character of the modes that cross into the corotation band varies. Only in one
exceptional case, for the simplest rotation law examined, do we observe a
mode crossing the corotation band boundary and
continuing to exist as a regular stable mode.  We think that such regular
modes within the corotation band are unlikely to exist for more realistic rotation laws. 
In general, at all points where modes
cross the corotation band boundary we observe the development of zero-step
 solutions.  At some
points we also observe the development of dynamically unstable modes.  A
necessary (but not sufficient) condition for dynamical instability is that
the equilibrium vorticity have a turning point on the shell.  Some of the
rotation laws examined in Paper I meet this condition when the degree of
differential rotation exceeds a certain threshold value.  All modes crossing the corotation
band boundary above this threshold become dynamically unstable.  In addition,
we observe for one rotation law the development of dynamical instability
via the merger of two stable modes outside the corotation band.  

Paper I raised a number of questions.  Is it possible to derive a necessary
and sufficient condition for instability? Do the singular
solutions of the continuous spectrum have physical meaning?  And do the
zero-step solutions appear distinct from the 
continuous spectrum in time
evolutions?  In this paper we resolve these questions by addressing the
initial value problem for the shell. By trying to isolate the various contributions
to the evolution of smooth initial data, we hope to clarify the physical 
role of different parts of the spectrum of a differentially rotating system.
The results for this simple toy model will guide us in future investigations of
the three dimensional problem \citep{wat03b}. 

\section{Key results from the mode analysis}
\label{key}
In Paper I we analysed the normal mode problem for the
differentially rotating spherical shell.  Since we will refer to a number of the
results from the mode analysis in this paper, we summarise the key
results in this section.  Consider an incompressible fluid on a differentially rotating spherical shell
(with radius $R$).  Assuming that the perturbations are $\propto
\exp(i m\varphi)$ we have the $\theta$-component of the perturbed Euler equation, 

\begin{equation}
\label{thetaeuler}
\partial_t \delta v_\theta + im\Omega \delta v_\theta -
2\Omega\cos\theta\delta v_\varphi = -\frac{1}{\rho R} \partial_\theta
\delta P,
\end{equation}
and the $\varphi$-component
\begin{equation}
\label{phieuler}
\partial_t \delta v_\varphi + im\Omega \delta v_\varphi + \tilde{\Omega} \delta
v_\theta = -\frac{im}{\rho R \sin\theta} \delta P,
\end{equation}
where the equilibrium vorticity is
\begin{equation}
\tilde{\Omega} = 2\Omega \cos\theta + \sin\theta \partial_\theta \Omega.
\end{equation}
The perturbed continuity equation for an incompressible fluid is 
\begin{equation}
\partial_\theta (\sin\theta \delta v_\theta) + im\delta v_\varphi = 0.
\end{equation}
This tells us that only toroidal modes are permitted in this problem.  
Defining the stream function (in the standard way)
\begin{equation}
U = \delta v_\theta \sin\theta,
\end{equation}
we can combine equations (\ref{thetaeuler}) and (\ref{phieuler}) to give
the vorticity equation
\begin{equation}
\label{vorthetat}
(i\partial_t - m\Omega) \nabla_\theta^2 U + \frac{m\partial_\theta
\tilde{\Omega}}{\sin\theta} U = 0,
\end{equation}
where 
\begin{equation}
\nabla^2_\theta = \partial_\theta^2 +
\frac{\cos\theta}{\sin\theta}\partial_\theta - \frac{m^2}{\sin^2\theta}
\end{equation}
is the Laplacian on the unit sphere. Changing variables to $x = \cos\theta$ the
vorticity equation becomes 
\begin{equation}
\label{voryt}
(i\partial_t - m\Omega) \nabla^2_x U - m \tilde{\Omega}' U = 0,
\end{equation}
where 
\begin{equation}
\nabla^2_x = (1-x^2)\partial_x^2 - 2x\partial_x -
\frac{m^2}{1-x^2},
\end{equation}
and we use a prime to denote derivatives with respect to $x$.  In terms
of $x$ the equilibrium vorticity is 
\begin{equation}
\tilde{\Omega} = 2x\Omega - (1-x^2) \Omega'.
\end{equation}
Assuming that we are interested in a mode solution behaving as $\exp(-i
\omega t)$, where $\omega$ is the frequency, equation (\ref{voryt}) becomes

\begin{equation}
(\omega-m\Omega) \nabla^2_x U -  m \tilde{\Omega}' U  = 0. 
\label{vorteq3}
\end{equation}
At corotation points ($x = x_c$), $\omega = m\Omega(x_c)$ and the equations
are formally singular. The nature of the solutions in the vicinity
of the corotation point can be analysed using the method of Frobenius.  The
general solution is found to be

\begin{equation}
U = U_\mathrm{reg} + U_\mathrm{sing},
\label{ugen}
\end{equation}
where

\begin{equation}
U_\mathrm{reg} = \sum_{n=0}^\infty a_n
(x-x_c)^{n+1},
\label{ureg}
\end{equation}
and

\begin{eqnarray}
U_\mathrm{sing}  =  \sum_{n=0}^\infty
& \bigg [& a_n (x-x_c)^{n+1} \ln (x-x_c) {}\nonumber \\ &&  + c_n (x-x_c)^n
\bigg] 
~~~~~~~~~~~~~~~~~x>x_c  {}\nonumber 
\\
U_\mathrm{sing}  = 
 \sum_{n=0}^\infty
& \bigg[ & b_n (x_c/x - 1)^{n+1} \ln
(x_c/x - 1)  {}\nonumber \\ && + d_n(x_c/x - 1)^n  \bigg] 
~~~~~~~~~~~~~~x<x_c.
\label{using}
\end{eqnarray}
We use separate solutions for $U_\mathrm{sing}$ on either side of the corotation
point to avoid taking the
logarithm of a negative quantity.  The two solutions for $U_\mathrm{sing}$ are matched such that $U$ is
continuous at the corotation point.  The first derivatives
are however discontinuous at $x_c$, with (in general) both a logarithmic
singularity and a step discontinuity.  The size of this step discontinuity
in $U'$ varies as one traverses the frequency band spanned by the
continuous spectrum, and for certain special frequencies it vanishes. We
refer to the solutions at these frequencies as zero-step solutions. At
the zero-step frequencies the Wronskian of the two linearly independent
solutions to equation (\ref{vorteq3}) vanishes,  despite the
logarithmic discontinuity in the first derivative.  The importance of
this will become apparent.

Let us now consider dynamical instabilities.  From equation
(\ref{voryt}) it is possible to derive the necessary (but not sufficient) condition for
instability $\tilde{\Omega}' = 0$.  
This condition is met for some of the rotation laws considered in Paper I
when the degree of differential rotation exceeds
some threshold value.  Dynamical instabilities are then found to
develop both where modes meet the corotation band boundary, and where modes
merge outside the corotation band.

\section{Evolving initial data: Green's function approach}
\label{green}

In this paper we consider equation (\ref{voryt}) as an initial-value problem.
That is, we are interested in the evolution of prescribed initial 
data, say, 
\begin{equation}
U(x, t=0)= U_0(x).
\end{equation} 
We first transform the problem into the frequency domain via the 
Laplace transform (we use $s\to -i\omega$ in order to allow direct comparison
with the mode calculations in Paper I).
\begin{equation}
\hat{U} = \int_0^\infty U e^{i\omega t} dt,
\end{equation}
with inverse 
\begin{equation}
U = \frac{ 1}{2 \pi} \int_{-\infty+ic}^{\infty+ic} \hat{U} e^{-i\omega t} 
d\omega. 
\label{lapinv}
\end{equation}
Applying this transform to equation (\ref{voryt}) we find 
\begin{equation}
(\omega - m\Omega) \nabla^2_x \hat{U} -  m \tilde{\Omega}^\prime
\hat{U} = i \nabla^2_x U_0.
\label{gov}
\end{equation}
Note that $\nabla^2_x U_0$ is always non-zero for this problem.  The only
non-zero smooth initial data for which $\nabla^2_x U_0 = 0$ would violate the boundary
conditions.    We should also make a brief comment on the 
Laplace transform.  For growing oscillatory
$U$, with time dependence $ \exp[(\alpha + 
i\beta)t]$, equation (\ref{gov}) is only correct if the imaginary part of 
$\omega$, $\omega_I$, is greater than $\alpha$. Otherwise the transform
would fail to converge as $t\rightarrow \infty$.  We must therefore ensure
that the constant 
$c$, which appears in the limits of integration of the 
inverse Laplace transform in equation (\ref{lapinv}), lies above all of the
poles of the integrand.  This means that we must have
$c > \alpha$.  Thus on the line integral, we always have $\omega_I > \alpha$,
and we never breach the assumption that we made when taking the Laplace
transform. We then assume analytic continuation of the Laplace transform into the
entire complex $\omega$ plane, and use the residue theorem to evaluate the
line integral. This does not change the situation: the line
integral itself always has $\omega_I > \alpha$ and the Laplace transform
and its inverse are well-defined.

We will now use the Green's function method to derive a formal solution 
to this equation (assuming for the moment that the problem is regular).
Given a solution $G(\omega,x,y)$ to 
\begin{equation}
\nabla^2_x G -  \frac{ m \tilde{\Omega}^\prime }{\omega - m\Omega} G  = \delta(x-y),
\label{geq}
\end{equation}
we have
\begin{equation}
\hat{U}(\omega, x) = i  \int_0^1 \frac{  G(\omega, x,y)}{\omega -m\Omega(y)}   \nabla^2_y
U_0 (y) dy.
\label{geqsol}
\end{equation}
Note that because of the symmetry of the problem we need only consider the upper 
hemisphere of the shell.
Finally, inverting the transform we obtain the solution to the 
initial value problem
\begin{equation}
U(x,t) = \frac{ i}{2\pi} \int_{-\infty+ic}^{ \infty+ic}d\omega~  e^{-i\omega t} 
\int_0^1   \frac{ G(\omega, x,y)}{\omega -m\Omega(y)}  \nabla^2_y
U_0 (y) dy.
\label{refe}
\end{equation}
The analytic properties of the integrand determine the spectrum
of the problem.

In the standard way, we construct the required Green's function $G(\omega,
x, y)$ from 
solutions to the homogeneous problem. Thus we want to solve
\begin{equation}
(\omega - m\Omega) \nabla^2_x \hat{U} - m  \tilde{\Omega}^\prime \hat{U} = 0. 
\label{ode}
\end{equation}
Consider two solutions, $\hat{U}_L$ and $\hat{U}_R$,  that 
satisfy the boundary conditions at the left and right endpoints
of the interval, respectively. Using the differential equation one can 
easily show that the Wronskian of any two solutions can be written
\begin{equation}
W = \hat{U}_L\hat{U}_R^\prime- \hat{U}_R\hat{U}_L^\prime =  \frac{
\tilde{W} (\omega)}{1-x^2}.
\label{wronskian}
\end{equation}
Pick an arbitrary point $y \in [0,1]$.
Then an overall solution, that is continuous 
at $x=y$, can be written
\begin{equation}
{\cal U}  = C \left[ H(x-y) \hat{U}_L(y) \hat{U}_R(x) 
+ H(y-x) \hat{U}_R(y) \hat{U}_L(x) \right].
\label{solcalu}
\end{equation}
Working out the derivatives of this function, and using equation (\ref{ode}),
 we find that ${\cal U}$ is a solution to
\begin{equation}
\nabla^2_x {\cal U} -   \frac{ m\tilde{\Omega}^\prime}{\omega - m \Omega} 
{\cal U} = \delta (x-y) C \tilde{W}(\omega).
\label{ode2}\end{equation}
Comparing this to the equation for the Green's function in the previous section
we see that we can identify
\begin{eqnarray}
G (\omega, x, y)  = \frac{ {\cal U}(\omega, x, y)}{C \tilde{W}(\omega)} &  =&
\frac{ 1 }{\tilde{W}(\omega)}  \Big[ H(x-y) \hat{U}_L(y)
\hat{U}_R(x){}\nonumber \\ &&  + 
H(y-x) \hat{U}_R(y) \hat{U}_L(x) \Big], 
\label{time2}
\end{eqnarray}
(clearly independent of the normalisation of $\hat{U}_L$ and $\hat{U}_R$).

\section{Solving the initial value problem}
\label{analytic}

To solve the initial value problem analytically we would need to evaluate
the integrals:

\begin{equation}
U(x,t) =   \frac{ i}{2\pi} \int_{-\infty+ic}^{ \infty+ic} 
\frac{{\cal J}(x,y)  e^{-i\omega t}}{C \tilde{W}}d\omega
\label{I1}
\end{equation}

\begin{equation}
{\cal J}(x,y)= \int_0^1   \frac{{\cal U} (\omega,x,y)  \nabla^2_y
U_0 (y)}{[\omega - m\Omega(y)]} dy, 
\label{I2}
\end{equation}
where ${\cal U}$ is given by equation (\ref{solcalu}).  Unfortunately, we
cannot in general solve even the homogeneous problem (for $\hat{U}_L$ and 
$\hat{U}_R$) analytically.  It is however possible to generate specific toy
problems for which the initial value problem may be solved analytically.
We are particularly interested in the behaviour of the continuous spectrum
and the zero-step solutions.  We therefore develop two toy problems.  The
first is a
particular rotation law that admits only a continuous 
spectrum as a solution.  An analytic solution of the initial value problem
is possible for simple initial data, and the solution sheds light on the
physical nature of the perturbation.  In the second problem we aim
to investigate the nature of the zero-step solutions. Ideally one would like to find
a simple rotation law, with a zero-step 
solution, for which one could solve the initial value problem analytically.
We have not been able to identify such a law.  Instead we have developed a
toy mathematical problem with solutions that include both a continuous spectrum
and a zero-step solution.  The behaviour of the eigenfunctions of this
system near the critical point is mathematically identical to the behaviour
that we find for the shell problem.  The initial value problem for our toy
system can however be solved analytically, guiding us towards a
quasi-analytic solution
for the zero-step solutions and continuous spectrum of the shell problem.  
The details of the two toy problems are contained in Appendices
\ref{simple} and \ref{notsosimple}.  In this section we summarise only the
key results from each problem.  

Appendix \ref{simple} presents a full analytic solution of the initial
value problem on the shell,
for a simple rotation law that possesses only a continuous spectrum (with
no zero-step solutions).  For simple initial data the collective perturbation associated
with the continuous spectrum is found to be non-singular, and hence
physical.  The perturbation dies away with time, with inverse power law
decay rather
than exponential decay.  The frequency of the decaying perturbation depends
in part upon position.  Similar behaviour has been found for the continuous
spectrum of a differentially rotating cylinder \citep{bal84b}.

Appendix \ref{notsosimple} introduces a toy problem that has
singular
continuous spectrum eigenfunctions with behaviour similar to those of our
problem.  By careful selection of
boundary conditions, one can ensure the presence of a single 
zero-step solution within the continuous spectrum. In this case we
are able to make some progress towards an analytic solution of the initial
value problem.  We draw several important conclusions that are likely to transfer
to the differentially rotating shell.  Firstly, the physical perturbation
contains a constant amplitude oscillatory component at the zero-step
frequency.  Secondly, just as in Appendix \ref{simple}, the rest of the
continuous spectrum dies away with 
time for simple initial data.  Again we find inverse power law decay rather than
exponential decay.  The frequency of the perturbation associated with the
continuous spectrum is, as before, found to depend in part upon position.  

Let us now proceed as far as we can with the analytic solution to the initial
value problem for the differentially rotating shell.  
We start by considering the integral
over position, equation (\ref{I2}).  We will write $\omega - m\Omega(y) \equiv
\bar{\Omega}(\omega,y) (y-y_c)$, where the corotation point $y_c$ is a function
of $\omega$.  Equation (\ref{I2}) becomes

\begin{eqnarray}
{\cal J}(x,y) & = & \hat{U}_L(x) \int_x^1 \frac{\hat{U}_R(y) \nabla^2_yU_0
(y)}{\bar{\Omega}(\omega,y) (y-y_c)} dy {}\nonumber \\ && + \hat{U}_R(x) \int_0^x
\frac{\hat{U}_L(y) \nabla^2_y U_0
(y)}{\bar{\Omega}(\omega,y) (y-y_c)} dy.
\label{pvreq}
\end{eqnarray}
If $\omega$ is in the range of $m\Omega$, then $0<y_c<1$ and the integrand
will be singular at $y=y_c$.  In Paper I we examined the variation of $\hat{U}$ with $y$ and found that
in the vicinity of the corotation point ($y\approx y_c$), for real $\omega$, it behaves as 

\begin{equation}
\hat{U} = \sum_{n=0}^\infty c_n (y-y_c)^n + \sum_{n=0}^\infty  a_n (y - y_c)^{n+1} \ln(y-y_c).
\label{hatu1}
\end{equation}
Whether $y_c$ is greater than or less than $x$ will depend on
$\omega$.  If $y_c > x$, the first integral in equation (\ref{pvreq}) has a
pole and a branch point at $y = y_c$, and we must treat the integral as a
principal value integral.  If on the other hand $y_c < x$, then it is the second
integral that must be treated as a principal value integral.  The same procedure is followed in Appendix
\ref{notsosimple}.   These integrals
are not easily evaluated analytically; one must in 
general resort to numerical methods to solve even the homogeneous problem
for $\hat{U}_L$ and $\hat{U}_R$.
However, by examining the power series
expansions around the singular point, which are
identical to the power series expansions around the singular point for the
problem in Appendix \ref{notsosimple}, it is easy
to see that the principal value 
integrals must result in smooth, non-singular functions.  Thus even without
solving the integrals analytically, we can state that the first integral
will give rise to a function $I_1(x,\omega)$ and the second integral to a
function $I_2(x,\omega)$, where both  $I_1$ and $I_2$ are smooth,
non-singular functions.  The frequency integral, equation (\ref{I1}), becomes

\begin{eqnarray}
U(x,t) & = &   \frac{ i}{2\pi} \int_{-\infty+ic}^{ \infty+ic} \frac{1}{\tilde{W}(\omega) }
 \Big[ \hat{U}_L(\omega,x) I_1(x,\omega)  {}\nonumber \\ && + \hat{U}_R(\omega,x) I_2(x,\omega)\Big]
 e^{-i\omega t} d\omega.
\end{eqnarray}
We cannot solve this integral analytically.  Instead, we will use Cauchy's
theorem to extract its properties.  We complete the contour using a
semicircle of infinite radius in the lower half plane that must be deformed
around any branch cuts.  The contribution from the semicircle integral vanishes because
the integrand tends to zero on this part of the contour.  The only
non-zero contribution will come from the poles and branch cuts 
of the integrand.
 
We know from Paper I that there are zeroes of $\tilde{W}$ at frequencies
at which there are discrete normal modes, and at
frequencies within the continuous spectrum where there are zero-step
solutions.  $\tilde{W}$ has a zero of multiplicity one in all cases except
where modes or zero-step solutions merge: at these frequencies it has a
zero of multiplicity two.  

We must also consider the behaviour of $\hat{U}$.  This function is
well-behaved and non-singular except for $\omega$ within the range of $m\Omega$. Let us
first consider the behaviour near the frequency $\omega = m\Omega(x)$, the
corotation frequency.  We need to know how $\hat{U}$ varies
with $\omega$ for fixed $x$.  This would in principle require a numerical 
calculation, but for the fact that we do know how
$\hat{U}$ varies with $x$ for fixed $\omega$ in the vicinity of the point
$x=x_c$.  Because $x_c$ is a
function of $\omega$, we can rewrite our power series expansions as
an expansion around the singular frequency rather than the singular
position (see Appendix \ref{notsosimple}, where the equivalence is
clearer).  We find that $\hat{U}$ is non-singular but has a branch point at
$\omega = m\Omega(x)$.  The two end points of the continuous spectrum, $\omega =
m\Omega_{\mathrm{min}}$ and $\omega = m\Omega_{\mathrm{max}}$ are also branch
points. The contour of integration should look similar to that
shown in Figure \ref{xsep}.   

We can now make some definitive statements about the
time-dependence of the physical perturbation $U(x,t)$.  
\begin{enumerate}
\item{
The poles of the integrand at discrete
frequencies $\omega_n$  
where the Wronskian is zero will
contribute terms with simple oscillatory time dependence $\exp(i\omega_n
t)$.  This is true both for discrete normal modes and the zero-step
solutions. Thus
by virtue of having zero Wronskian, the zero-step frequencies should appear
virtually indistinguishable from the standard discrete normal modes,
despite being in the continuous spectrum.}
\item{
At frequencies where modes merge
and the Wronskian has a double pole we will have additional terms of the
form $t\exp(i\omega_n t)$.  Such linear growth signals the onset of
instability, as can
be seen if one considers an unstable mode exhibiting exponential growth $\exp[(i\omega_n
+\omega_m)t]$.  For small $\omega_m$ this can be expressed as a Taylor
series:

\begin{equation}
e^{i\omega_n t} e^{\omega_m t} \approx e^{i\omega_n t}\left[ 1 + \omega_m t
+ \dots\right].
\end{equation}
This behaviour is to be expected
since mode mergers usually lead to dynamical instability \citep{sch80}.}
\item{
The behaviour of the continuous spectrum is more difficult to ascertain but,
since the end points of the corotation band on the real line are 
branch points, we would expect decaying oscillatory behaviour similar to
that found in Appendix \ref{notsosimple}. The frequency of
oscillation will inevitably depend in part upon position.   We cannot rule out the
possibility of constant amplitude oscillation, or even growth of the continuous
spectrum, in response to some
initial data.  What is clear, however, is that the
perturbation associated with the continuous spectrum is  
non-singular and hence physical.}
\end{enumerate}

\section{A sufficient instability condition}
\label{instab}

Having discussed the general properties of the solution, we will
now use an analytic approach to the 
problem to derive a sufficient condition for
instability.  We follow the method outlined by \citet{bal99} for inviscid
plane parallel shear flow.   We start by reference to equation (\ref{ode2}), the equation obeyed by
${\cal U}$.  Let us assume that we can normalise the solution in such
a way that 

\begin{equation}
\int_0^1 \nabla_x^2 {\cal U} dx = \Lambda.
\label{norm}
\end{equation}
We assume that $\Lambda$ is real (note that, strictly speaking, $\Lambda$ will depend on $\omega$ and
$y$). Then we find from equation (\ref{ode2}) that
\begin{equation}
\Lambda = C \tilde{W} +  \int_0^1 \frac{ m \tilde{\Omega}^\prime}{ 
\omega - m \Omega} {\cal U} dx.
\end{equation}
Hence
\begin{equation}
\epsilon \equiv C\tilde{W} = \Lambda - \int_0^1 \frac{
m\tilde{\Omega}^\prime}{\omega - m\Omega} {\cal U} dx.
\end{equation}

In the limit when $\omega$ is real but outside the range $m\Omega(x_c)$
for $x_c$ in [0,1], $\epsilon$ is real. $\epsilon$ has zeroes
at frequencies where there are normal mode solutions outside the corotation
band.
If on the other hand $\omega$ is real, but in the range  $\omega = m
\Omega(x_c)$ for $x_c \in [0,1]$,  then the integrand is singular and the
integral must be interpreted as the principal value\footnote{If the
eigenfunction ${\cal U}$ were regular at the corotation point then the
integrand would not be singular.  This applies to both types of regular
eigenfunction described in Paper I, and can be seen from the power series
expansions of the regular eigenfunctions around the singular point.}.
\begin{eqnarray}
\epsilon \equiv C \tilde{W} & = & \Lambda - {\cal P} 
\int_0^1 \frac{ m \tilde{\Omega}^\prime }{\omega - m \Omega} {\cal U}(\omega,x,y) dx
{}\nonumber \\ && \pm i \pi \frac{ \tilde{\Omega}^\prime (x_c) }{
\Omega^\prime(x_c)} {\cal U}(\omega, x_c,y). 
\label{epsdef}
\end{eqnarray}
The sign depends on whether we circle the singularity in the
upper or the lower half-plane.  This quantity  $\epsilon$, as defined in equation (\ref{epsdef}),
can be used to derive a sufficient instability criterion.  Our next step is to derive a
useful expression for ${\cal U}$.   

In Section (\ref{green}) we constructed a Green's function 
for the initial value problem, equation (\ref{time2}), that depended on ${\cal
U}$, which depends in turn on the
solutions, $\hat{U}$, to the homogeneous problem.  The equation for the
homogeneous problem, equation (\ref{ode}) may be
rewritten as

\begin{equation}
\nabla^2_x \hat{U} =  \frac{ m\tilde{\Omega}^\prime}{ \omega - m\Omega
}\hat{U}.  
\end{equation}
We can solve this equation using the Green's function $G_0(x,x^\prime)$
that satisfies

\begin{equation}
\nabla^2_x G_0 = \delta(x-x^\prime).
\label{gdel}
\end{equation} 
The explicit form for this Green's function 
is given in Appendix \ref{simple}. Finally, we have

\begin{equation}
\hat{U}(x) = \int_0^1  \frac{ m \tilde{\Omega}^\prime  }{ \omega -
m\Omega }  \hat{U} G_0(x,x^\prime)dx^\prime.
\label{uint} 
\end{equation}
Knowing that we can find $G_0$, we can rewrite equation (\ref{ode2}) for
${\cal U}$ as an integral 
equation. We obtain

\begin{eqnarray}
{\cal U} (\omega,x,y) & = & C \tilde{W} G_0(x,y){}\nonumber \\ && + 
\int_0^1 \frac{m\tilde{\Omega}^\prime}{\omega - m \Omega} 
{\cal U}(\omega, x^\prime,y) G_0(x,x^\prime) dx^\prime, 
\label{homog}
\end{eqnarray}
or, if we use the normalisation, equation (\ref{norm}),

\begin{eqnarray}
{\cal U} (\omega,x,y)  & = &   \Lambda G_0(x,y)  + 
 \int_0^1 \frac{ m\tilde{\Omega}^\prime}{\omega - m \Omega} 
{\cal U}(\omega,x^\prime,y){}\nonumber \\ &&  [ G_0(x,x^\prime) - G_0(x,y) ] dx^\prime. 
\label{bm1}
\end{eqnarray}
Recall that we have so far assumed that $\omega$ is complex in order to avoid 
singularities on the real $x$-axis. If we relax this assumption, we 
need to allow for the possibility that $\omega - m \Omega(x_c) = 0$.
Then we find that (\ref{bm1}) is a singular problem. However, we can regularise
this equation by a judicious choice of $y$. We therefore take $y=x_c$ in (\ref{bm1})
to get

\begin{eqnarray}
{\cal U} (\omega,x,x_c) & = & \Lambda G_0(x,x_c) + 
\int_0^1 \frac{ m\tilde{\Omega}^\prime}{ \omega - m \Omega} 
{\cal U}(\omega,x^\prime,x_c) {}\nonumber \\ && [ G_0(x,x^\prime) - G_0(x,x_c) ] dx^\prime. 
\label{bm2}
\end{eqnarray}
As discussed by \citet{bal99}, a
solution to the Fredholm equation (\ref{bm2}) is a singular eigenfunction 
associated with a continuous spectrum frequency. 
Note that we can let ${\cal U} (\omega,x,x_c)\to {\cal U} (\omega,x)$ 
since $x_c$ is determined by $\omega$ anyway.  

Equations (\ref{epsdef}) and (\ref{bm2}) hold for all points where the
normalisation given by equation (\ref{norm}) is valid.   There are however certain solutions within
the continuous spectrum for which 
this normalisation is not appropriate.  If we have solutions for which
$\tilde{W} = 0$ then equation (\ref{ode2}) for ${\cal U}$ is identical to the normal mode
equation, (\ref{vorteq3}), that we solved in Paper I.  Thus for zero
Wronskian solutions we can identify ${\cal U}$ with the normal mode
solutions found in Paper I.  Most zero
Wronskian solutions correspond to modes with finite and continuous
eigenfunctions outside the corotation band.  The zero-step solutions are an
exception, having zero Wronskian and a logarithmic discontinuity in the first derivative.
So for the frequencies where zero-step solutions exist we must treat the
integral in
equation (\ref{norm}) as a principal value, with 

\begin{equation}
\Lambda = \int_0^1 \nabla_x^2 {\cal U} dx = {\cal P}\int_0^1\nabla_x^2 {\cal U} dx
\pm i \pi\left[(x-x_c)\nabla_x^2 {\cal U}\right]_{x=x_c}.
\label{norm2}
\end{equation}  
For these special frequencies, 

\begin{eqnarray}
\epsilon & = & {\cal P}\int_0^1\left[\nabla_x^2 {\cal U} - 
\frac{ m\tilde{\Omega}^\prime }{\omega - m \Omega} {\cal U}\right] dx
 {}\nonumber \\ &&  \pm  i \pi \left[\left[(x-x_c)\nabla_x^2 {\cal U}\right]_{x_c}
  +  \frac{\tilde{\Omega}^\prime(x_c)}{\Omega^\prime(x_c)} {\cal U}(\omega,
x_c,y)\right].
\label{epsdef2}
\end{eqnarray}
That the real part of $\epsilon$ should be zero if ${\cal U}$
is a solution to the normal mode problem is clear from
equation (\ref{vorteq3}).  That the imaginary part must also be zero can be seen by
taking the limit as $x\rightarrow x_c$ of $(x-x_c)$ times equation
(\ref{vorteq3}), and applying
L'H\^opital's rule. So as we would expect, $\epsilon$ (and hence
$\tilde{W}$) is zero at frequencies
where there are zero-step solutions to the normal mode problem.
For all other frequencies within the continuous spectrum $\epsilon$ will in general be
complex and non-zero.  

At frequencies where there are zero-step solutions, the regularising
term in the integral in equation (\ref{bm2}) cancels exactly the term
involving $\Lambda$, and equation
(\ref{bm2}) becomes

\begin{equation}
{\cal U} (\omega,x) =  
\int_0^1 \frac{ m\tilde{\Omega}^\prime}{\omega - m \Omega} 
{\cal U}(\omega,x^\prime) G_0(x,x^\prime) dx^\prime. 
\label{bm3}
\end{equation}
The equation becomes homogeneous, with a kernel that is truly
singular.  The presence of homogeneous solutions at certain frequencies
 complicates any attempts to solve equation (\ref{bm2}) directly.
Fortunately, we do not need to solve this equation to derive a sufficient
instability criterion.  

Let us recap.  Given a solution to the integral equation (\ref{bm2}), we can
find the quantity $\epsilon$ defined in equation (\ref{epsdef}). 
These two equations hold for all frequencies in the
continuous spectrum apart from those for which there are zero-step
continuous spectrum solutions to the normal mode problem.  
At these frequencies, $\epsilon$ will
be zero and there will be a homogeneous solution to equation (\ref{bm2}),
as discussed above. 

Let us rescale equations (\ref{bm2}) and (\ref{epsdef}) according to ${\cal
U}/\Lambda \to {\cal U} $ and 
$\epsilon/\Lambda \to \epsilon$ to get
\begin{eqnarray}
{\cal U} (\omega,x) & =  & G_0(x,x_c) + 
\int_0^1 \frac{ m\tilde{\Omega}^\prime}{ \omega - m \Omega} 
{\cal U}(\omega,x^\prime){}\nonumber \\ &&  [ G_0(x,x^\prime) - G_0(x,x_c) ] dx^\prime, 
\label{bm2s}
\end{eqnarray}
and

\begin{equation}
\epsilon   =  1 - {\cal P} 
\int_0^1 \frac{m\tilde{\Omega}^\prime}{\omega - m \Omega} {\cal U} 
(\omega,x)dx
 \pm i \pi \frac{\tilde{\Omega}^\prime (x_c)}{\Omega^\prime(x_c)}
{\cal U}(\omega, x_c).
\label{epsdefs}
\end{equation}
In order to have an instability we must have a zero of the
Wronskian $\tilde{W}$ in the upper half of the $\omega$-plane.
The Nyquist method tells us that the number of zeroes ${\cal N}$ of
$\tilde{W}$ in a region enclosed by a contour 
$\cal{C}$ in the $\omega$ plane is given by the integral

\begin{equation}
{\cal N} = \frac{1}{2\pi i}
\int_{\cal C}\frac{\tilde{W}^\prime}{\tilde{W}}d\omega =
\frac{1}{2\pi} \Delta \mathrm{arg}(\tilde{W}),
\label{wind}
\end{equation}
where $\Delta$ is used to denote the change in a quantity.  The integral in
equation (\ref{wind})
tells us the number of times that the function 
$\tilde{W}$ winds around the origin of the complex $\tilde{W}$ plane
as $\omega$ varies around $\cal C$.  We decompose the integral
into two parts:

\begin{equation}
\frac{1}{2\pi i}
\int_{\cal{C}}\frac{\tilde{W}^\prime}{\tilde{W}}  d\omega = \frac{1}{2\pi i}
\int_{\cal{R}}\frac{\tilde{W}^\prime}{\tilde{W}}  d\omega + \frac{1}{2\pi i}
\int_{\cal{C}'}\frac{\tilde{W}^\prime}{\tilde{W}} d\omega.
\label{fullint}
\end{equation}
The first integral is taken over the real line $\cal{R}$, the second
over a semicircle ${\cal C}'$ in the upper half plane.  In Appendix \ref{semic} we
show that  the integral over the semicircle vanishes. Now $\epsilon(\omega) = 
C(\omega) \tilde{W}(\omega)$, so

\begin{equation}
\int_{\cal R} \frac{\tilde{W}^\prime}{\tilde{W}} d\omega = \int_{\cal R}
\frac{\epsilon^\prime}{\epsilon} d\omega
- \int_{\cal R} \frac{C^\prime}{C} d\omega.  
\label{ints}
\end{equation}
$C$ was an arbitrary constant in equation (\ref{solcalu}), so we can
choose $C$ to be real.  Then, setting $\epsilon =
|\epsilon|\exp(i\mathrm{arg} \epsilon )$ and $\tilde{W}  = |\tilde{W}|
\exp(i\mathrm{arg} \tilde{W})$, equation (\ref{wind}) becomes

\begin{equation}
{\cal N} = \frac{1}{2\pi} \Delta \mathrm{arg} (\epsilon).
\end{equation}
In other words, the number of unstable eigenvalues
should be determined by the change in the argument of $\epsilon$ as
$\omega$ traverses the real line.  In fact, we can constrain the range of 
integration still further.  For
frequencies outside the corotation band, the principal value integral in
equation (\ref{epsdefs}) becomes a regular integral with no imaginary part
(since we can normalise ${\cal U}$ to ensure that it is real). Hence, $\epsilon$ is
real for all frequencies outside the range $m\Omega_{\mathrm{min}} <
\omega < m\Omega_{\mathrm{max}}$ and cannot wind.  The presence of a
corotation band and the associated continuous spectrum is clearly
essential for instability on the shell.  Thus

\begin{equation}
{\cal N} = \frac{1}{2\pi} \Delta \mathrm{arg} (\epsilon) =
\int_{m\Omega_{\mathrm{min}}}^{m\Omega_{\mathrm{max}}}
\frac{\epsilon^\prime}{\epsilon} d\omega.  
\end{equation}

The existence of $\epsilon = 0$ solutions at certain frequencies
within the corotation band, and for certain degrees of differential rotation
on the edges of the corotation band, will not affect the winding
number. This is because these are 
isolated points where $\epsilon$ jumps directly to zero; and there is no
winding if $\epsilon$ passes through the origin rather than around it.  The
frequencies where $\epsilon 
= 0$ can be excluded from the 
calculation of winding number \citep{bal95}, and will not be discussed
further. 

At the endpoints of the corotation band, $\omega = m\Omega_{\mathrm{min}}$
and $\omega = m\Omega_{\mathrm{max}}$, one can demonstrate that $\epsilon =
1$.  This means that there can only be winding if we cross the negative real $\epsilon$-axis
while traversing the band.  

From equation (\ref{epsdefs}), it is clear that
the imaginary part of $\epsilon$ is zero if
either ${\cal U}(\omega, x_c) = 0$ or $\tilde{\Omega}^\prime(x_c) =
0$.  The first condition,  ${\cal U}(\omega, x_c) = 0$, cannot lead to instability,
because if this is the case we have a regular mode within the
corotation band, and the real part of $\epsilon$ would also vanish.  Thus
for winding we require 

\begin{equation}
\tilde{\Omega}^\prime(x_c) = 0,
\end{equation}
which is consistent with the necessary criterion for instability derived
in Section \ref{instab}. We denote the frequency at which this inflexion point condition is met as
$\omega_i$.  

For there to be at least one unstable eigenvalue, the real
part of $\epsilon$ must be negative at the frequency $\omega_i$. We therefore have a
sufficient condition for instability

\begin{equation}
{\cal{P}} \int_0^1 \frac{ m\tilde{\Omega}^\prime }{\omega_i - m
\Omega} {\cal U}(\omega_i, x) dx > 1.
\end{equation}
The above condition tells us that there is at least one unstable
eigenvalue.  But what if we want to know the absolute number of unstable
eigenvalues?  In Paper I we found the total number of unstable eigenvalues
by computing normal modes.  The method outlined in this section offers in
principle an alternative.   If we could solve equation (\ref{bm2s})
numerically, using a Fredholm equation solving routine, we could use the
resulting ${\cal U}$ to compute $\epsilon$ as the frequency varies across
the continuous spectrum range, and hence determine the winding
number directly.  A similar calculation has been carried out by \citet{bal99} for
plane parallel flow.  Their problem is however simpler as there are no
homogeneous solutions to their equivalent of equation (\ref{bm2s}).  In our
case one would have to begin by solving the homogeneous (eigenvalue)
Fredholm problem, as stated in equation (\ref{bm3}), and then exclude these
frequencies from the main calculation.  We have assessed
the feasibility of
solving the homogeneous equation using the numerical methods for singular
integral equations discussed in \citet{del85}, but conclude that
application of these methods
is far from trivial in this case.  Computation of normal modes is a far
more convenient method of determining the total number of unstable
eigenvalues for this problem.  

\section{Evolving initial data: Numerical results}
\label{numerics}

The prime way to check the mode calculations is the numerical
evolution of generic smooth
initial data.  This will also enable us to check the quasi-analytic
solution to the initial value problem formulated in Section~\ref{analytic}.  

We start by reformulating equation (\ref{vorthetat}) to
ensure that our variables are real valued. We introduce
\begin{equation}
U(t,\theta,\varphi) = C(t,\theta)\cos m \varphi - S(t,\theta) \sin m\varphi
\end{equation}
After decoupling the sine/cosine parts, we arrive at two equations that 
can be  written succinctly as 
\begin{eqnarray}
\partial_t A - m\Omega B + m \partial_\theta \tilde{\Omega} S &=& 0
\label{pta} \\
\partial_t B + m\Omega A - m \partial_\theta \tilde{\Omega} C &=& 0
\label{ptb}
\end{eqnarray}
where $A$ and $B$ are defined by
\begin{eqnarray}
\nabla^2_\theta C &=& \frac{ A}{\sin \theta}  \label{ABA} \\ 
\nabla^2_\theta S &=& \frac{ B}{\sin \theta} \label{ABB}
\end{eqnarray}
In order to get a first-order system we introduce
\begin{eqnarray}
Z& =& \partial_{\theta}C
\label{ZC} \\
X& =& \partial_{\theta}S
\label{XS}
\end{eqnarray}
so equations (\ref{ABA}) and (\ref{ABB}) become
\begin{eqnarray}
\sin\theta A-\sin^2\theta\partial_{\theta}Z-\sin\theta\cos\theta Z+m^2C& =&
0
\label{AZC} \\
\sin\theta B-\sin^2\theta\partial_{\theta}X-\sin\theta\cos\theta X+m^2S & =
& 0
\label{BXS}
\end{eqnarray}
We now have six first order equations for six unknowns. We can see from
equations (\ref{AZC})-(\ref{BXS}) that for regularity at the poles we
require $C=S=0$.

Our numerical solution of the six equations 
is based on a second order accurate
Crank-Nicholson implicit finite differencing scheme.
The finite differenced equations are solved using a relaxation
scheme with $601$ gridpoints and a timestep $\Delta t = 0.05
\Delta \theta$.  Convergence testing has verified that the code is second order
accurate, and stability analysis indicates that the code is stable
over many dynamical timescales.  To test the code we
 ran time evolutions for the simple  
rotation law described in Appendix \ref{simple}. The analytical
solution of the initial value problem was used to verify the robustness of the
numerical time evolution code.  We found the results to agree extremely
well with those shown in Figure \ref{ntwo}. 

We then performed time evolutions for a more
complicated rotation law, the solar rotation law that we studied in Paper
I (Law 3 of that paper):

\begin{equation}
\Omega = 2\pi\left(454.8 - 60.4\beta \cos^2\theta - 71.4\beta \cos^4
\theta \mathrm{~~nHz} \right)
\end{equation}
The parameter $\beta$ measures the degree of differential rotation,
uniform rotation corresponding to $\beta = 0$.  

\begin{figure}
\centering
\includegraphics[height=6cm,clip]{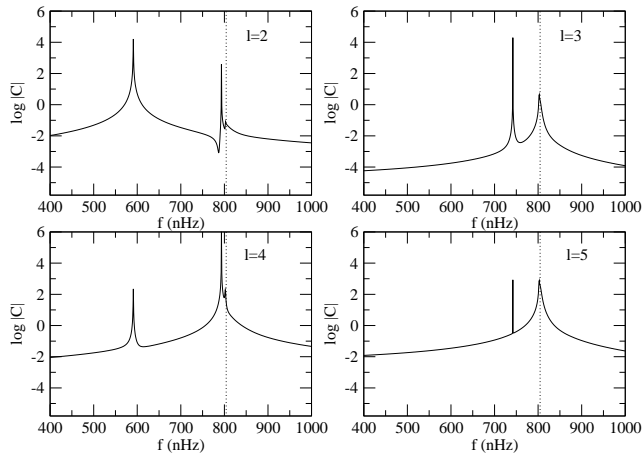}
\caption{Power spectrum of numerical time evolution of data for the solar
differential rotation law with $m=2$ and $\beta=0.4$. The four 
panels represent initial data proportional to the spherical harmonic $Y^2_l$, with $l=2-5$. 
The sampling position is $\theta
= \pi/30$.  The dashed line marks the
lower edge of the corotation band, at 804.6.  We know from Paper I
that at $\beta = 0.4$ the $l$ = 2, 3, and 4 modes are outside the
corotation band,
at frequencies 592.1, 744.3 and 796.0 respectively.  Peaks at these
frequencies are clearly visible.  The $l=5$ mode does not cross
into the corotation band but the mode results predict a zero-step solution just
inside corotation, at frequency 806.35.  A peak at this point is  visible in
the power spectrum.}
\label{m2b0.4}
\end{figure}

\begin{figure}
\centering
\includegraphics[height=6cm,clip]{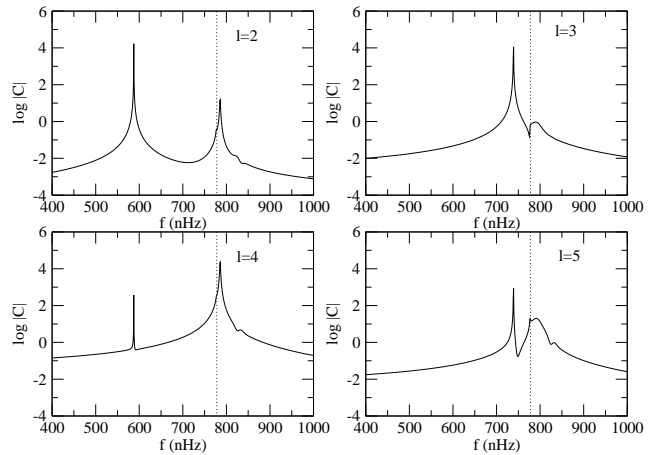}
\caption{Power spectrum of numerical time evolution of data for the solar
differential rotation law with $m=2$ and $\beta=0.5$.
The four 
panels represent initial data proportional to the spherical harmonic $Y^2_l$, with $l=2-5$. 
The sampling position is $\theta
= \pi/30$. The dashed line marks the
lower edge of the corotation band, at 777.8.  We know from Paper I
that at $\beta = 0.5$ the $l$ = 2 and 3 modes are outside the
corotation band,
at frequencies 588.66 and 741.08 respectively.  Peaks at these
frequencies are clearly visible.  The $l$ = 4 mode is now inside the
corotation band, at frequency 788.34.  This is visible, in addition to a
zero-step solution excited by $l=5$ initial data at 802.50.
The peak at around 825 for $l$ = 4 and 5 initial data is probably 
associated with the continuous spectrum and is an artefact of sampling
position.}
\label{m2b0.5}
\end{figure}

For initial data we used the standard spherical
harmonics $Y_l^m$ that are the eigenfunctions of $U$ in uniform rotation.  A sample
of numerical results for this rotation law with $m=2$ are shown in
Figures~\ref{m2b0.4}-\ref{m2b0.5}.  For frequencies outside the
corotation band,
there are clear peaks at the  frequencies of the regular normal modes.  
For dynamically unstable modes we are able to
verify the growth times predicted by the normal mode
analysis.  What is most interesting, however, is the behaviour of the
continuous spectrum. The power spectrum shows distinct 
peaks within the continuous spectrum at
frequencies where we found zero-step solutions in Paper I, just as
predicted in Section \ref{analytic}.  Although the
size of the peak in the power spectrum varies depending on 
sampling position (the 
value of $x$ at which
we monitor the time evolution), these frequencies clearly stand out from the
rest of the continuous spectrum. 

We were also interested to see how the rest of the continuous spectrum
behaved in response to an initial perturbation.  The continuous spectrum was observed to die
away with time for smooth initial data, and to be excited to a much
lower extent than the modes and zero-step solutions.

The numerical evolutions support the conclusions from our Green's function 
analysis. In particular, it is clear that the zero-step solutions stand 
out from the rest of the continuous spectrum. Hence, it is reasonable to 
expect them to be of particular relevance for the dynamical 
evolution. This would be in accordance with the results of
\citet{bal85}, which show that the merger of zero-step solutions
may lead to dynamical instabilities.  

\section{Discussion}

The main question arising from the normal mode analysis of Paper I
was the nature of the physical perturbations associated with the singular
eigenfunctions of both the zero-step corotating solutions, and the
continuous spectrum within which they exist.  By studying the initial value problem both
analytically and numerically we have been able to resolve this issue.   

Firstly, we have confirmed that the physical perturbations associated
with these phenomena are not singular.  Their time dependence in
response to an initial perturbation, however, is complicated.  For
appropriate initial data, the zero-step solutions behave in much the
same way as stable normal modes outside corotation. That is,
they produce a clear peak in the power spectrum at a fixed frequency,
standing out from the rest of the continuous spectrum.  The fact that 
such oscillatory
solutions can exist within the continuous spectrum 
is of great relevance for future
studies of the oscillations of differentially rotating systems.  

The behaviour of the rest of the continuous spectrum is difficult to
predict analytically, and can only be elucidated by numerical time
evolutions.  The analytic calculations suggest that the physical
perturbation associated with the continuous spectrum is oscillatory, with
a complicated frequency dependence that varies in part according to
the position on the shell.  The amplitude of the oscillations could
grow, remain at constant amplitude or die away with time (as a an
inverse power of time), depending on the initial data and the rotation law.  For the
rotation laws and initial data that we have studied we observe only
decay, but we cannot rule out either constant amplitudes or growth.  
  
The results of numerical time evolutions have verified to high
accuracy the predictions of mode frequencies made by the normal mode
analysis.  This is true both for modes outside the corotation band (including
dynamically unstable modes), and the
zero-step solutions within it.  This is encouraging, as when we move
on to study more realistic stellar models, we are likely to rely more
on time evolutions than on increasingly complicated mode analysis. 
 
\section{Acknowledgments}

We would like to thank Bernard Schutz and Horst Beyer for useful discussions.
We acknowledge support from the EU 
Programme `Improving the Human Research Potential and the Socio-Economic
Knowledge Base' (Research Training Network Contract HPRN-CT-2000-00137).
ALW is supported by a PPARC postgraduate studentship and NA acknowledges
support from the Leverhulme Trust in the form of a prize fellowship.

\appendix

\section{A rotation law that exhibits
only a continuous spectrum}
\label{simple}

Solving the initial value problem analytically for
an arbitrary rotation law is extremely difficult.  For certain rotation
laws, however, an analytical solution 
of the initial value problem is both feasible and instructive.   In 
this Appendix we analyse a rotation law that exhibits a
continuous spectrum but no other type of oscillation.  Our aim is to
isolate the response of the continuous spectrum to an initial
perturbation.  We select the simple rotation law

\begin{equation}
\Omega = \frac{\Omega_c}{1+|x|},
\end{equation}
for which $\tilde{\Omega}' = 0$. Solving the normal mode problem for
this rotation law, one finds that the only solutions
form a continuous spectrum with derivatives that are
discontinuous at the corotation point.  In contrast to the general case,
however, the derivatives possess only a step discontinuity at the
corotation point (no logarithmic singularity).  There are no zero-step
solutions.  

\begin{figure}
\centering
\includegraphics[width=8cm, height = 4cm, clip]{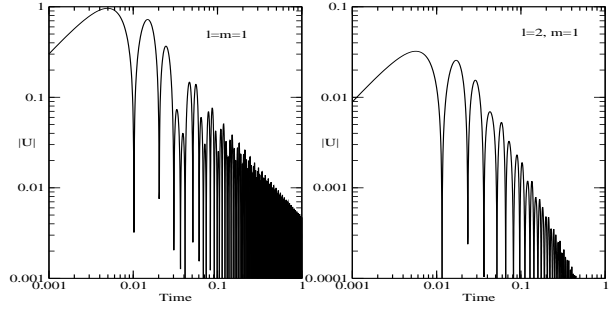}
\caption{Time evolution of $|U|$ for $l=m=1$ and $l=2,m=1$ initial data.  At
late times, $|U|$ falls off as $1/t$ for the  $l=m=1$ data and as  $1/t^2$ 
for the $l=2,m=1$ data. } 
\label{ntwo} 
\end{figure}

Returning to the initial value problem, equation (\ref{gov}) becomes
\begin{equation}
\nabla_x^2 \hat {U} = \frac{i\nabla_x^2 \hat{U}_0}{(\omega - m\Omega)}.
\end{equation}
We solve this equation using a Green's function $G_0(x,y)$ that
satisfies

\begin{equation}
\nabla_x^2 G_0 = \delta(x - y).
\end{equation}
Because of the symmetry of the governing equation there are two
possible boundary conditions at $x=0$, either $dG_0/dx(0,y)=0$ (even
solutions), or $G_0(0,y) = 0$ (odd
solutions). For even solutions, we find that

\begin{equation}
G_0(x,y)= \left\{\begin{array}{ll} 
\frac{-1}{2m} \left[\left(\frac{y+1}{y - 1}\right)^{\frac{m}{2}} +
\left(\frac{y-1}{y+ 1}\right)^{\frac{m}{2}} \right]
\left(\frac{x-1}{x+1}\right)^{\frac{m}{2}}  & x>y \\
\frac{-1}{2m} \left[\left(\frac{x+1}{x-1}\right)^{\frac{m}{2}} +
\left(\frac{x-1}{x+1}\right)^{\frac{m}{2}} \right]
\left(\frac{y-1}{y+1}\right)^{\frac{m}{2}}     & x<y,\end{array}\right.
\label{greef1}
\end{equation}
and for odd solutions,

\begin{equation}
G_0(x,y)= \left\{\begin{array}{ll} 
\frac{1}{2m}\left[\left(\frac{y-1}{y+1}\right)^{\frac{m}{2}} -
\left(\frac{y+1}{y-1}\right)^{\frac{m}{2}}
\right]\left(\frac{x-1}{x+1}\right)^{\frac{m}{2}} & x>y \\
\frac{1}{2m}\left[\left(\frac{x-1}{x+1}\right)^{\frac{m}{2}} - \left(\frac{x+1}{x-1}\right)^{\frac{m}{2}}
 \right] \left(\frac{y-1}{y+1}\right)^{\frac{m}{2}} & x<y.\end{array}\right.
\label{greef2}
\end{equation}
We can now recover the physical perturbation using the inversion integral
\begin{equation}
U(x,t) = \int_0^1   {\cal H} (x,y)  \nabla^2_y
U_0 (y) dy, 
\label{inttime}
\end{equation}
where
\begin{equation}
{\cal H}(x,y) =   \frac{ 1}{2\pi} \int_{-\infty+ic}^{ \infty+ic} 
\frac{ G_0(\omega,x,y) e^{-i\omega t}}{
\omega -m\Omega (y)  }d\omega.
\label{time1}
\end{equation}
We illustrate the typical behaviour by giving results for the particular
case $U_0(y) = P_l^m(y)$, with $l=m$ (even).  In this case, $\nabla_y^2
U_0(y) \propto (1-y^2)^{m/2}$  and we find 

\begin{eqnarray}
\label{lm}
U(x,t) & = & \frac{i(-1)^{m+1}}{4\pi m} \bigg\{
\left(\frac{1-x}{1+x}\right)^{\frac{m}{2}}\int_0^x
\Big[(1+y)^m + {}\nonumber \\ && (1-y)^m \Big] e^{-im\Omega(y)t}  dy  +
\bigg[\left(\frac{1+x}{1-x}\right)^{\frac{m}{2}} +  {}\nonumber \\ &&
\left(\frac{1-x}{1+x}\right)^{\frac{m}{2}}\bigg]\int_x^1
(1-y)^m  e^{-im\Omega(y)t}dy  \bigg\}.
\end{eqnarray}
Evaluating the integral analytically for $l=m=1$ we find 
\begin{eqnarray} 
U(x,t) & = & 2b  \left[it\mathrm{Ei}(i\Omega_c t) - e^{-i\Omega_c
t}\right] 
+ \frac{(a+b)}{2}  {}\nonumber \\ && \left[(4 + 2it)e^{-\frac{i\Omega_c t}{2}} + (t^2 - 4it)
\mathrm{Ei}\left(\frac{i\Omega_c t}{2}\right)\right] 
{}\nonumber \\ && 
+ \frac{1}{2} \Big[-3a + b + 2(b-a)x - i(a+b)(1+x)t {}\nonumber \\ && + (a+b)x^2 
\Big]e^{-\frac{i\Omega_c t}{1+x}}
+ {}\nonumber \\ && \frac{1}{2}\left[4iat -  (a+b)
t^2\right]\mathrm{Ei}\left(\frac{i\Omega_ct}{1+x}\right), 
\end{eqnarray} 
where

\begin{equation} 
a(x) = \left(\frac{1+x}{1-x}\right)^{\frac{1}{2}},~~ 
b(x) = \left(\frac{1-x}{1+x}\right)^{\frac{1}{2}}, 
\end{equation} 
and the asymptotic expansion of the exponential integral $\mathrm{Ei}(x)$ is
 
\begin{equation}
\mathrm{Ei}(x) = \frac{e^{-x}}{x}\left(1-\frac{1}{x} + \frac{2}{x^2} -
...\right).
\end{equation}
Expanding the exponential integrals we find that all terms whose amplitudes
 grow with $t$ cancel.  The resulting time dependence is illustrated
in Figure \ref{ntwo}.  The most persistent term has an amplitude that decays as
$1/t$. In addition,
the three different frequency components $\Omega_c$, $\Omega_c/2$
and $\Omega_c/(1+x)$ interact to produce 
 beating.  The beating
is particularly pronounced if $x$ is chosen
such that $\Omega_c/(1+x)$ is very close to either of the other two
frequencies.

If instead we consider initial data $U_0(y) = P_l^m(y)$ with $l=m+1$ (odd) then $
 \nabla_y^2 U_0(y) \propto y(1-y^2)^{m/2}$ and we find that 

\begin{eqnarray}
\label{lmp1}
U (x,t) &= & \frac{i(-1)^m}{4\pi
m} \bigg\{ \left(\frac{1-x}{1+x}\right)^{\frac{m}{2}} \int_0^x y
\Big[(1-y)^m -  {}\nonumber \\ &&  (1+y)^m
\Big] e^{-im\Omega(y) t}  dy  +
\bigg[\left(\frac{1-x}{1+x}\right)^{\frac{m}{2}} - {}\nonumber \\ &&
\left(\frac{1+x}{1-x}\right)^{\frac{m}{2}} 
 \bigg] \int_x^1 y (1-y)^m
e^{-im\Omega(y)t} dy \bigg\}.
\end{eqnarray}
Evaluating the integral analytically, we obtain similar results to
those for $l=m$ initial data, except that now the most persistent terms decay as
$1/t^2$.  This is illustrated in Figure \ref{ntwo}.  

\section{A toy problem with a 
continuous spectrum and a zero-step solution} 
\label{notsosimple} 

In Appendix \ref{simple} we solved the initial value problem for the
differentially rotating shell for a rotation law that exhibits only a
continuous spectrum.  The continuous spectrum eigenfunctions obtained for
this rotation law differ from those found for more general rotation laws,
however, in two ways.  Firstly, the eigenfunctions do not possess a
logarithmic singularity in the first derivatives, only a finite step.  Secondly, there are no
zero-step solutions.   

We have not been able to identify a rotation law that, in addition to
exhibiting both of these features, is simple enough to permit analytical
solution of the initial value problem.  In this Appendix we instead analyse
a toy problem that has solutions with the required behaviour, to inform our
investigation of the shell problem.  

\subsection{Selection of a toy problem}

Consider the equation

\begin{equation} 
(x-i\partial_t) U''(x) + U(x) = 0, 
\label{toytdep}
\end{equation} 
where primes indicate derivatives with respect to $x$. Assume a normal
mode solution, with $U\propto \exp(-i\omega t)$.  Equation
(\ref{toytdep}) becomes 

\begin{equation} 
(x-\omega)U'' + U = 0. 
\label{toymode}
\end{equation} 
The critical point $x=\omega$, which gives rise to the continuous spectrum,
plays the same role as the corotation point on the differentially rotating shell.
The general solution to equation (\ref{toymode}) is  
 
\begin{equation} 
U = \sqrt{x - \omega}[A_1 J_1(2\sqrt{x - \omega}) + A_2 Y_1(2\sqrt{x - 
\omega})], 
\end{equation} 
where $J_\nu$ and $Y_\nu$ are Bessel functions of the first and second kind 
respectively, $\nu$ being the index of the function.  $A_1$ and $A_2$ are
constants.  To show that
the solutions have the required behaviour,
it is convenient to consider separate solutions on
either side of the critical point (compare to Paper I):

\begin{equation} 
U(x) = \left\{ \begin{array}{ll} U_<(x)  & x < \omega \\ U_>(x)  &
x > \omega, \end{array} \right. 
\end{equation} 
where
\begin{eqnarray} 
U_<(x) & = &  \sqrt{x - \omega} \left[c_1 J_1(2\sqrt{x - \omega}) + c_2
Y_1(2\sqrt{x - \omega})\right]
 \\ U_>(x) &  = & \sqrt{x -
\omega} \left[c_3 J_1(2\sqrt{x - \omega}) + c_4 Y_1(2\sqrt{x - 
\omega})\right]
\end{eqnarray} 
where $c_1$ - $c_4$ are constants.  Near the critical point: 

\begin{eqnarray} 
U_< & = & -\frac{c_2}{\pi} + {\cal O}(x - \omega) {} \\
U_> &= & -\frac{c_4}{\pi} + {\cal O}(x - \omega). 
\end{eqnarray} 
For continuity of the function at the critical point we require $c_2 =
c_4$.  Let us now consider the behaviour of the first derivatives near
the critical point:
\begin{eqnarray} 
U_<'  & =  & c_1 + \frac{c_2}{\pi} [ \ln|x - \omega| + i\pi + 2\gamma ]+ {\cal 
O}(x - \omega) {} \\ 
U_>' &   = &  c_3 + \frac{c_2}{\pi} [ \ln|x - \omega| + 2\gamma ]+ {\cal 
O}(x - \omega), 
\end{eqnarray}
where $\gamma$ is Euler's constant. Thus at the critical point there will
be both a logarithmic singularity and a finite step in the first
derivative.  The degree of freedom associated with the finite step in the
first derivative allows
us to construct a continuous spectrum of solutions for $\omega$ in the
range of $x$. Mathematically the behaviour is identical to that of
 the continuous spectrum on the differentially rotating shell. The
Wronskian of the two solutions, $W$,  is   
 
\begin{equation} 
W = \frac{1}{\pi}[c_4 c_1 - c_2 c_3 + ic_2 c_4]. 
\end{equation} 
Since $c_2 = c_4$ we see that $W$ is zero if $c_1 - c_3 + i c_2 =
0$. This corresponds to the finite step in the first derivative
vanishing, indicating the presence of a zero-step solution. As we are
interested in the behaviour of zero-step solutions we will select a
domain and boundary conditions such that there is at least one
zero-step solution.  We select the domain $x = [0,1]$ and fix one of
the boundary conditions, $U_<(0) = 0$.  In this case there is a zero-step
solution for $\omega = 1/2$ if we insist on a second boundary
condition  $U_>(1) = \beta$, where

\begin{equation}
\beta = c_3 \lim_{\omega \rightarrow 1/2} \left[
\sqrt{1-\omega}\frac{[f_2 f_3 - f_1 f_4 + if_1 f_3]}{f_2 -
if_1}\right], 
\end{equation}
and 

\begin{eqnarray}
f_1(\omega) & = & J_1(2\sqrt{-\omega}) {} \\
f_2(\omega) & = & Y_1(2\sqrt{-\omega}) {} \\
f_3(\omega) & = & J_1(2\sqrt{1-\omega}) {} \\
f_4(\omega) & = & Y_1(2\sqrt{1-\omega}).
\end{eqnarray}
There are no other zero-step solutions within the continuous spectrum
 with $0<\omega<1$.  Investigation suggests
that there are no other mode solutions for this problem with these boundary
conditions with either complex $\omega$ or
real $\omega$ outside the range $0<\omega<1$.  

\subsection{Setting up the initial value problem}

In the previous section we established the boundary conditions
necessary to ensure that the toy problem exhibits both a continuous
spectrum and a zero-step solution at $\omega = 1/2$.  We now define   

\begin{equation}
\hat{U} = \int_0^\infty U e^{-i \omega t} dt,
\end{equation}
and apply a Laplace transform to 
equation (\ref{toytdep}) to obtain the inhomogeneous equation

\begin{equation}
\hat{U}'' + \frac{\hat{U}}{x-\omega} = -\frac{iU_0''}{x-\omega}.
\label{gt1}
\end{equation}
We will solve this problem using a Green's function that obeys the equation

\begin{equation}
G'' + \frac{G}{x-\omega} = \delta(x-y).
\label{gt2}
\end{equation}

\subsection{Finding the Green's function}

We construct the Green's function as follows.  Let $\hat{U}_L$ and $\hat{U}_R$ be
solutions to the homogeneous equation that obey the boundary conditions on
the left and right of the domain respectively. The Wronskian of these
two solutions, $W$, is a function of frequency only.  Consider the function

\begin{equation}
F(x,y) = C[H(y - x) \hat{U}_R(y) \hat{U}_L(x) + H(x - y) \hat{U}_L(y)\hat{U}_R(x)].
\end{equation}
$F$ is continuous at the point $x=y$, irrespective of the
normalisation of $\hat{U}_L$ and $\hat{U}_R$.  It can be shown to obey the equation

\begin{equation}
F'' + \frac{F}{x - \omega} = C W(\omega) \delta(x-y).
\end{equation}
We can see that a valid solution to equation (\ref{gt2}) is

\begin{eqnarray}
G(\omega, x, y) & = & \frac{1}{W(\omega)} \Big[H(y - x) \hat{U}_R(y) \hat{U}_L(x)
{}\nonumber \\ && + H(x -
y) \hat{U}_L(y)\hat{U}_R(x)\Big]. 
\end{eqnarray}
We now need to determine $\hat{U}_L$
and $\hat{U}_R$.  From our examination of the normal mode problem, we know
that normalised solutions to the homogeneous problem are

\begin{equation}
\hat{U}_L(x) = \sqrt{x - \omega} \left[J_1(2\sqrt{x - \omega}) + \bar{c}_2 Y_1(2\sqrt{x -
\omega})\right]
\end{equation}

\begin{equation}
\hat{U}_R(x) = \sqrt{x - \omega} \left[J_1(2\sqrt{x - \omega}) + \bar{c}_4 Y_1(2\sqrt{x -
\omega})\right],
\end{equation}
where (comparing to the normal mode problem) we have defined $\bar{c}_2 =
c_2/c_1$, $\bar{c}_4 = c_4/c_3$, and $\bar{\beta} = \beta/c_3$.  The
Wronskian of the two solutions is  

\begin{equation}
W = \bar{c}_4 - \bar{c}_2 + i\bar{c}_2\bar{c}_4.
\end{equation}
The boundary condition at $x=0$, $\hat{U}_L(0) = 0$, gives

\begin{equation}
\bar{c}_2 = -\frac{J_1(2\sqrt{-\omega})}{Y_1(2\sqrt{-\omega})}.
\end{equation}
The boundary condition at $x=1$, $\hat{U}_R(1) =\bar{\beta}$, gives

\begin{equation}
\bar{c}_4 =
\frac{1}{Y_1(2\sqrt{1-\omega})}\left[\frac{\bar{\beta}}{\sqrt{1-\omega}} -
J_1 (2\sqrt{1-\omega}) \right].
\end{equation}
It can be verified that the Wronskian is indeed zero at  $\omega =
1/2$.  We have now found our Green's function.  

\subsection{The inversion integrals (part I)}

The first stage in recovering $U(x,t)$ is to use the
Green's function to find $\hat{U}(x,\omega)$.  Using equations
(\ref{gt1}) and (\ref{gt2}) we can show, in the standard way, that 

\begin{equation}
\hat{U}(x) = -i \int_0^1 \frac{U_0''G}{y-\omega} dy - \left[G \hat{U}' -\hat{U}
G'\right]_0^1.
\label{grc}
\end{equation}
The last term in equation (\ref{grc}) is non-zero because of our
choice of boundary conditions.  By choosing
 conditions that ensured the
presence of a zero-step solution, we have also ensured that $G$ and $\hat{U}$ do
not obey the same boundary conditions.  In the following sections we will
 neglect this second term because our primary interest is in
the zero-step solutions. Their contribution is encapsulated in the
position integral, not in the second term, which depends solely on the
end-points.  Restricting our attention to the position integral,

\begin{eqnarray}
\int_0^1 \frac{U_0''(y)G(\omega, x, y)}{y-\omega} dy & = & 
\frac{1}{W(\omega)}\Big[\hat{U}_L(x) \left(I_{p1} +  \bar{c}_4 (\omega) 
I_{p2}\right) {} \nonumber \\ && + \hat{U}_R(x)\left( I_{p3} + \bar{c}_2 (\omega) I_{p4}\right)\Big]
\end{eqnarray}
where

\begin{eqnarray}
I_{p1} & = & \int_x^1 \frac{J_1(2\sqrt{y - \omega})}{\sqrt{y -
\omega}}dy {}\\
I_{p2} & = &  \int_x^1 \frac{Y_1(2\sqrt{y - \omega})}{\sqrt{y -
\omega}}dy {} \\
I_{p3} & = & \int_0^x \frac{J_1(2\sqrt{y - \omega})}{\sqrt{y -
\omega}}dy {} \\
I_{p4} & = & \int_0^x \frac{Y_1(2\sqrt{y - \omega})}{\sqrt{y -
\omega}}dy.
\end{eqnarray}
In order to evaluate these integrals we must select initial data that obeys
the boundary conditions.  We will
consider the simple case 

\begin{equation}
U_0 (y)  =  y^2/2 + (\bar{\beta} - 1/2)y  ~~\rightarrow~~ U_0''  =  1.  
\end{equation}
$I_{p1}$ and $I_{p3}$ are regular at the point $y = \omega$, so the
integrals are straightforward: 

\begin{equation}
I_{p1} =  J_0(2\sqrt{x-\omega}) - J_0(2\sqrt{1-\omega}), 
\end{equation}

\begin{equation}
I_{p3} =  - J_0(2\sqrt{x-\omega}) + J_0(2\sqrt{-\omega}).
\end{equation}
The integrals $I_{p2}$ and $I_{p4}$ require a little more care because the
integrands are singular at the point $y = \omega$ (real $\omega$ only). If $x< \omega< 1$, we
must treat $I_{p2}$ as a principal value integral, and $I_{p4}$ is
non-singular.  If on the other hand $0< \omega< x$, we
must treat $I_{p4}$ as a principal value integral, and $I_{p2}$ is
non-singular. Thus we get

\begin{equation}
I_{p2} =  Y_0(2\sqrt{x-\omega}) -
Y_0(2\sqrt{1-\omega}) - H(\omega - x) i, 
\end{equation}

\begin{equation}
I_{p4} =  - Y_0(2\sqrt{x-\omega}) +
Y_0(2\sqrt{-\omega}) - H(x-\omega) i. 
\end{equation}

\subsection{The inversion integrals (part II)}

\begin{figure}
\includegraphics[height = 5.5cm]{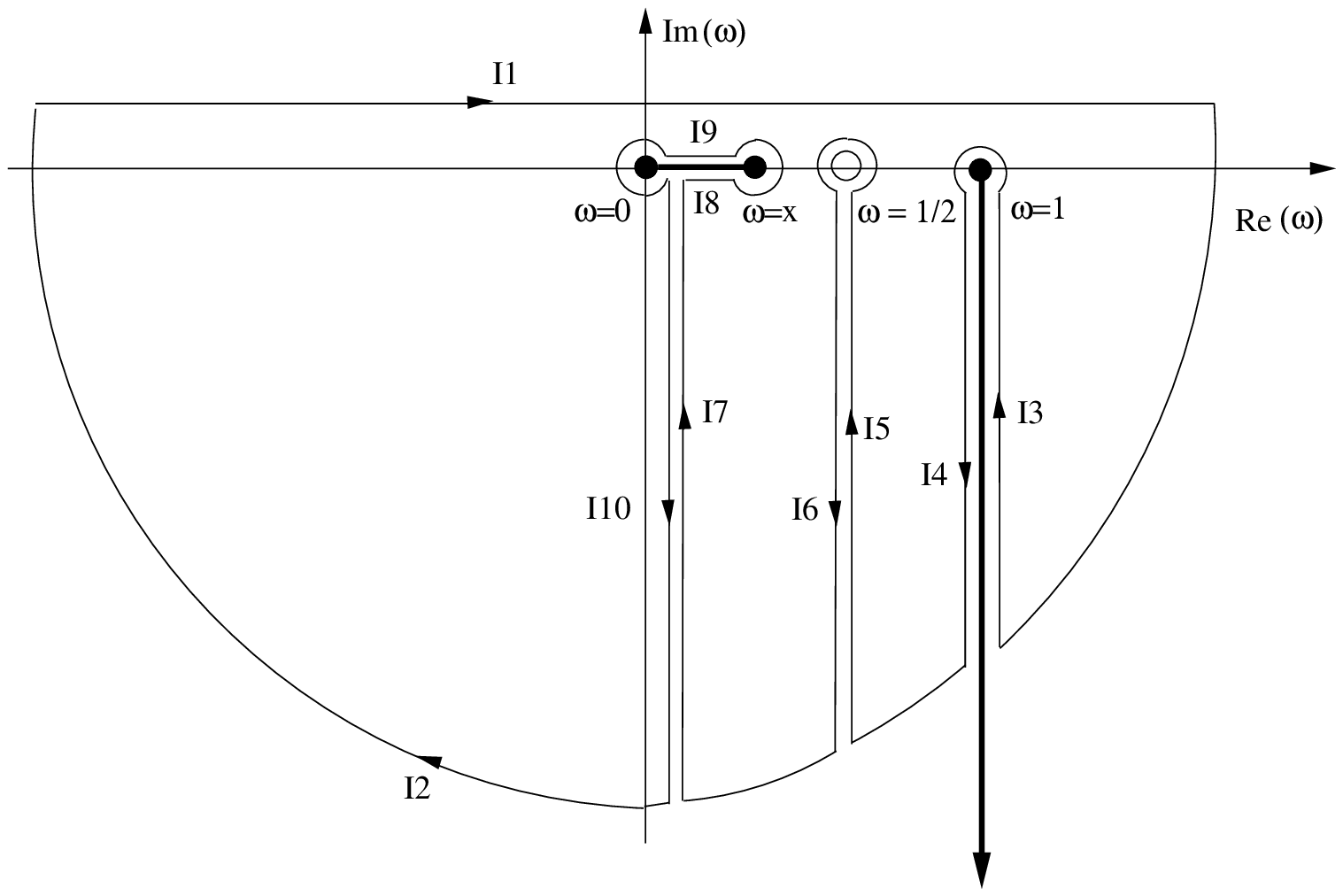}
\caption{The contour for the frequency inversion integral.  Filled dots
indicate branch points, open dots isolated poles.  The thick solid lines are
branch cuts.}
\label{xsep}
\end{figure}

The final inversion integral to recover the time dependent solution
$U(x,t)$ is 

\begin{equation}
U(x,t) = \frac{1}{2\pi} \int_{-\infty + ic}^{\infty + ic} \hat{U}(\omega,
x)e^{-i\omega t} d\omega.
\label{toyfc}
\end{equation}
For our given choice of initial data, equation (\ref{toyfc}) becomes:

\begin{eqnarray}
U(x,t) & = & \frac{1}{2\pi} \int_{-\infty + ic}^{\infty + ic} \frac{e^{-i\omega t}
\sqrt{x-\omega}}{W(\omega)}\Big\{J_1(2\sqrt{x-\omega})  {}\nonumber \\ && \left[J_0(2\sqrt{-\omega})
- J_0(2\sqrt{1-\omega})\right] + 
J_1(2\sqrt{x-\omega}) {}\nonumber \\ && \big[(\bar{c}_4 - \bar{c}_2)Y_0(2\sqrt{x-\omega}) -
\bar{c}_4 Y_0(2\sqrt{1-\omega}) + {}\nonumber \\ && \bar{c}_2 Y_0(2\sqrt{-\omega})\big] +
 Y_1(2\sqrt{x-\omega})\big[(\bar{c}_2 -
\bar{c}_4){}\nonumber \\ && J_0(2\sqrt{x-\omega}) + \bar{c}_4 J_0(2\sqrt{-\omega}) -
\bar{c}_2 J_0(2\sqrt{1-\omega})\big] {}\nonumber \\ &&
- i\bar{c}_2\bar{c}_4 Y_1(2\sqrt{x-\omega}) -
iJ_1(2\sqrt{x-\omega}){}\nonumber \\ && \left[ H(\omega - x)\bar{c}_4 +
H(x-\omega)\bar{c}_2\right]\Big\}.
\label{full1}
\end{eqnarray}
The integrand has a simple pole at  $\omega = 1/2$ due to the zero of the
Wronskian.  In
addition, there are branch points at $\omega = 0, x, 1$.  The contour of
integration for $x<1/2$ is shown in Figure 
\ref{xsep} (if $x>1/2$, we would instead join the points $\omega = x$ and $\omega = 1$ by a
 branch cut, and have an isolated branch cut down from $x=0$). The integral
around the semicircle, I2, vanishes as the radius tends to infinity, while
the integrals along I5/I6 and I7/I10 cancel exactly.    

The residue associated with the simple pole at $\omega= 1/2$ is $\propto
\exp(-it/2)$.  Such constant amplitude oscillatory frequency contributions are typical of
frequencies where there are zero-step solutions.  Identical behaviour
is found for the shell model, as discussed in
Section \ref{analytic}.

Working out the contributions from I3, I4, I8 and I9 is more 
difficult.  We will focus on I8/I9, as the continuous spectrum contribution
arises from the branch point at $\omega= x$.  We restrict our attention to
the case when $x$ is small (for illustrative purposes).  Making this
assumption we note that $\omega$ and $\omega - x$ are also small on I8
and I9. We will use Taylor expansions of the
integrand using the three small parameters $x$, $\omega$, and $\omega - x$ .  

Keeping only terms up to second order, we find that the
contribution to equation
(\ref{full1}) from the integrals along I8 and I9, $I_\omega$,  is

\begin{eqnarray}
I_\omega & = & \int_{I8, I9} e^{-i\omega t}(W_0 + W_1
\omega)\Big[ \alpha_1 (x-\omega)\ln(x-\omega)  {}\nonumber \\ &&
 + \alpha_2 (x-\omega)^2\ln(x-\omega)  + 
\alpha_3 \omega(x-\omega)\ln(x-\omega) {}\nonumber \\ &&  + \alpha_4 \omega^2\ln(-\omega) +
\alpha_5 \omega(x-\omega)\ln(-\omega)\Big], 
\end{eqnarray}
where $\alpha_1$ - $\alpha_5$ are constants, and we have expanded the inverse Wronskian in terms of
the small parameter $\omega$:  

\begin{equation}
W(\omega)^{-1} = W_0 + W_1\omega + {\cal O} (\omega^2),
\end{equation}
where $W_0$ and $W_1$ are also constants.  We now choose the phase of
the logarithms such that

\begin{equation}
\ln(-\omega) = \left\{ \begin{array}{ll} \ln|\omega| & \mathrm{~~~~On~I8}
\\
\ln|\omega| + 2\pi i  & \mathrm{~~~~On~I9}
\end{array}\right.
\end{equation}

\begin{equation}
\ln(x-\omega) = \left\{ \begin{array}{ll} \ln|x-\omega| & \mathrm{~~~~On~I8}
\\
\ln|x-\omega| + 2\pi i  & \mathrm{~~~~On~I9.}
\end{array}\right.
\end{equation}
The logarithm terms cancel when we integrate over I8 and I9, so the
only non-zero contribution will come from the $2\pi i$ piece on I9.  At
first order, the contribution to $I_\omega$ is

\begin{equation}
I_{\omega1} =  \beta_1 \left[\frac{1 - ixt - e^{-ixt}}{t^2}\right],
\end{equation}
where $\beta_1$ is a constant.  The second order contribution to $I_\omega$ is

\begin{equation}
I_{\omega2} = \frac{1}{t^3} \left[ \gamma_1  + \gamma_2 xt + \gamma_3 x^2
t^2 + \left(\delta_1 + \delta_2 xt + \delta_3 x^2t^2\right) e^{-ixt}\right],
\end{equation}
where $\gamma_1 - \gamma_3$ and $\delta_1 - \delta_3$ are
constants.  If we include higher order terms we find that that
the most persistent contribution to $U(x,t)$ arising from the
continuous spectrum decays as $1/t$.  Higher order terms merely introduce contributions
that decay as successively higher order powers of $t$.  It is also clear
that the continuous spectrum contributes a term to $U(x,t)$ with a
frequency that depends on position $x$.  

The analysis presented in this section is for a specific choice of initial
data and for small $x$.  We must therefore ask whether the key results
would change if we had made different choices. It is certainly by no means
clear that the continuous spectrum should decay for all smooth initial data.
Although we have not observed anything other than decay, we cannot rule out the possibility of
constant amplitude oscillation or even growth.  In contrast, the existence of a
frequency component that depends on position appears inevitable.

\section{Vanishing of semicircle integral in derivation of instability criterion}
\label{semic}
In this Appendix we will demonstrate that we can neglect the integral over
the semicircle in equation (\ref{fullint}) by considering the nature of
solutions for large $|\omega|$. We start by rewriting equation ($\ref{ode}$) as

\begin{equation}
\nabla^2_x \hat{U} - \frac{m}{\omega}\left[\Omega \nabla^2_x \hat{U} -
\tilde{\Omega}' \hat{U}\right] = 0.
\label{eqpar}
\end{equation}
We expand $\hat{U}$ in terms of the small parameter $m/\omega$

\begin{equation}
\hat{U} = \hat{U}_0 + \frac{m}{\omega}\hat{U}_1.
\end{equation}
Equation (\ref{eqpar}) becomes

\begin{eqnarray}
\nabla^2_x \hat{U}_0 + \frac{m}{\omega}\nabla^2_x \hat{U}_1  -
\frac{m}{\omega}\Big[\Omega \nabla^2_x \hat{U}_0 + \frac{m}{\omega}\Omega
\nabla^2_x \hat{U}_1 && {}\nonumber \\ -
\tilde{\Omega}' \hat{U}_0 - \frac{m}{\omega}\tilde{\Omega}'
\hat{U}_1\Big] & = & 0.
\label{eqpar2}
\end{eqnarray}
At leading order equation (\ref{eqpar2}) becomes

\begin{equation}
\nabla^2_x \hat{U}_0 = 0.
\label{u0}
\end{equation}
At first order equation (\ref{eqpar2}) becomes

\begin{equation}
\nabla^2_x \hat{U}_1 + \tilde{\Omega}' \hat{U}_0 = 0.
\label{u1}
\end{equation}
The general solution to equation (\ref{u0}) is
  
\begin{equation} 
\hat{U}_0 = \left\{\begin{array}{llll} \hat{U}_{0R} & = & c_1 \left(\frac{x-1}{x+1}\right)^{\frac{m}{2}} & 
x>y \\ \hat{U}_{0L} & = & c_2 \left(\frac{x+1}{x-1}\right)^{\frac{m}{2}} + c_3 
\left(\frac{x-1}{x+1}\right)^{\frac{m}{2}} & x<y, \end{array} \right.
\label{u0sol}
\end{equation}
where for continuity of the function at $x = y$ we require

\begin{equation}
c_1 = c_2\left(\frac{y + 1}{y - 1}\right)^m + c_3.
\end{equation}
It is clear from equation (\ref{u0sol}) that $\hat{U}_0$ does not depend on
$\omega$.  Using the method of Green's functions to solve equation (\ref{u1}) we must
first solve 

\begin{equation}
\nabla_x^2 G_0 = \delta(x-x').
\end{equation}
The solution to this equation is given in Appendix \ref{simple}, equations
(\ref{greef1}) and (\ref{greef2}).  We find
$G_0(x,x')$ to be independent of $\omega$.  We now have

\begin{equation} 
\hat{U}_1 = \left\{\begin{array}{lll} \hat{U}_{1R} =  -\int G_0(x,x')
\tilde{\Omega}'(x') \hat{U}_{0R}(x',y) dx' & 
x>y \\ \hat{U}_{1L}  =  -\int G_0(x,x')
\tilde{\Omega}'(x') \hat{U}_{0L}(x',y) dx' & x<y. \end{array} \right. 
\label{u1sol}
\end{equation}
It is clear from equation (\ref{u1sol}) that $\hat{U}_1$ must also be
independent of $\omega$.  Let us consider the implications of this for the
Wronskian $W$.  We get

\begin{eqnarray}
W  & = & \hat{U}_{0L} \hat{U}_{0R}' - \hat{U}_{0R}\hat{U}_{0L}' +
\frac{m}{\omega} \Big[\hat{U}_{1L}\hat{U}_{0R}' - 
\hat{U}_{0R} \hat{U}_{1L}' {}\nonumber \\ && + \hat{U}_{0L}\hat{U}_{1R}' - \hat{U}_{1R}
\hat{U}_{0L}' \Big] + \sim {\cal O} 
\left(\frac{m}{\omega}\right)^2.  
\end{eqnarray}   
This means that we can write $\tilde{W} = (1-x^2)W$ as 

\begin{equation}
\tilde{W} = f(x,y) + \frac{m}{\omega} g(x,y),
\end{equation}
and we obtain

\begin{equation}
\frac{\tilde{W}'}{\tilde{W}} = - \frac{m
g(x,y) /\omega^2}{f(x,y)  + m g(x,y)/\omega} \approx - \frac{m g(x,y)
f(x,y)}{\omega^2}.
\end{equation}
Integrating over frequency,

\begin{equation}
\int_{{\cal C}'}\frac{\tilde{W}'}{\tilde{W}} d\omega =  -\frac{m g(x,y)
f(x,y)}{\omega}.
\end{equation}
In the limit as $\omega \rightarrow \infty$ the integral clearly
tends to zero.  We can therefore neglect the integral 
over the infinite semicircle in equation (\ref{fullint}).  
\end{document}